\documentclass{PoS}

\def\pt  {p_{T}}

\def\gsim{\,\lower.25ex\hbox{$\scriptstyle\sim$}\kern-1.30ex%
\raise 0.55ex\hbox{$\scriptstyle >$}\,}
\def\lsim{\,\lower.25ex\hbox{$\scriptstyle\sim$}\kern-1.30ex%
\raise 0.55ex\hbox{$\scriptstyle <$}\,}

\newcommand{\gev}{\ensuremath{\mathrm{GeV}}}

\newcommand{\tev}{\ensuremath{\mathrm{TeV}}}

\newcommand{\MET}{\mbox{\ensuremath{E \kern-0.6em\slash_{\rm T}}}}
\newcommand{\MHT}{\mbox{\ensuremath{H \kern-0.75em\slash_{\rm T}}}}

\newcommand{\ra}{\ensuremath{\rightarrow}}

\newcommand{\fbi}{\ensuremath{\mathrm{fb}^{-1}}}

\title{Searches for leptoquark production and compositeness at the Tevatron}

\ShortTitle{Searches for leptoquark production and compositeness at the Tevatron}

\author{\speaker{Thomas Nunnemann}\thanks{for the D0 and CDF Collaborations}\\
        Ludwig-Maximillians Universit\"{a}t, Munich, Germany\\
        E-mail: \email{Thomas.Nunnemann@lmu.de}}

\abstract{Recent searches for leptoquark production and compositeness in 
$p\bar{p}$ collisions at $\sqrt{s}=1.96\,\tev$ are presented using data 
samples with integrated luminosities up to $4\,\fbi$ recorded with the 
D0 and CDF detectors at the Tevatron collider.}

\FullConference{European Physical Society Europhysics Conference on High Energy Physics,
EPS-HEP 2009,\\
		 July 16 - 22 2009\\
		 Krakow, Poland}

\begin{document}

\section{Search for leptoquark production}
Numerous extensions of the standard model (SM) predict the existence of
leptoquarks, i.e. colored bosons which carry both lepton and quark 
quantum numbers and thus allow lepton-quark transitions~\cite{lqpheno}.
At hadron colliders,
leptoquarks are predominantly produced in pairs via the strong coupling.
Single leptoquarks can be produced via $t$-channel leptoquark exchange,
which depends on the unknown leptoquark-lepton-quark coupling $\lambda$.

The pair-production of scalar leptoquarks is a pure QCD 
process (when
neglecting the contribution from $t$-channel lepton exchange which is 
$\propto \lambda^2$) and has been calculated to NLO \cite{Kramer:1997hh}.
Thus its cross-section depends on no additional model parameter except
the assumed leptoquark mass $M_{LQ}$.
In case of vector leptoquarks, the pair-production cross section is generally
much larger and additionally depends on unknown anomalous couplings.
Furthermore, the cross section has only been calculated at LO. 

Leptoquarks could, in principle, decay into any combination of a quark and a lepton,
but leptoquarks with masses as low as $\mathcal{O}(100\,\gev)$ are only allowed
to couple
to one generation of quarks and leptons, since they otherwise would generate
lepton number violation and sizable flavor-changing neutral currents.
The branching fractions of the leptoquark decays into a charged lepton and
a quark or a neutrino and a quark are determined by the respective
$LQ$-$\ell$-$q$ coupling. Thus, leptoquark pair-production can produce
three characteristic final states with rates determined by the branching 
fraction $\beta=\mathcal{B}(LQ\rightarrow \ell^\pm q)$: 
$\ell^+ q \ell^- q$, $\ell^\pm q \nu q$, and $\nu q \nu q$.

\subsection{Leptoquarks in the acoplanar jet topology}
Both the D0 and CDF collaborations searched for pair production
of scalar leptoquarks in event topologies with two acoplanar jets
and missing transverse momentum $\MET$ using data sets corresponding to
an integrated luminosity of 2.5\,\fbi{} and 2\,\fbi, 
respectively~\cite{Abazov:2008at,CDFexoticweb}. To discriminate the
leptoquark signal from the background, consisting mainly of 
$Z(\rightarrow\nu\nu)+$ jets and $W(\rightarrow\ell\nu)+$ jets production,
$\MET$ and the scalar sum of the jet transverse momenta, 
$H_{T}=\sum_{\mathrm{jets}}\pt$, were used as selection variables.
Upper limits on the cross section times branching ratio were obtained, defined
at 95\% C.L.,
and compared to the NLO prediction reduced by its uncertainty to derive
lower limits on the leptoquark mass.

Assuming $\beta = 0$, the CDF and D0 searches exclude scalar leptoquarks of
the first and second generation below 190\,GeV and 205\,GeV, respectively.

\subsection{First generation leptoquarks}
The D0 collaboration recently updated their search for the pair production
of first generation leptoquarks in $ejej$ and $ej\MET j$ final states using 
the Run\,IIa data set with an integrated luminosity of 
1\,\fbi~\cite{Abazov:2009gf}. For both channels, 
the main background contributions are associated production of vector bosons 
with jets and $t\bar{t}$ production. In the $ejej$ channel, the signal was 
discriminated using the dielectron invariant mass $M_{ee}$, the scalar sum
of the final state objects' transverse momenta $S_T$, and the average 
electron-jet
invariant mass, whilst in the $ej\MET j$ channel the selection is based on
$S_T$ and the final state objects' $p_T$.

\begin{figure}
\includegraphics[width=.50\textwidth,bb=6 -7 524 368,clip]{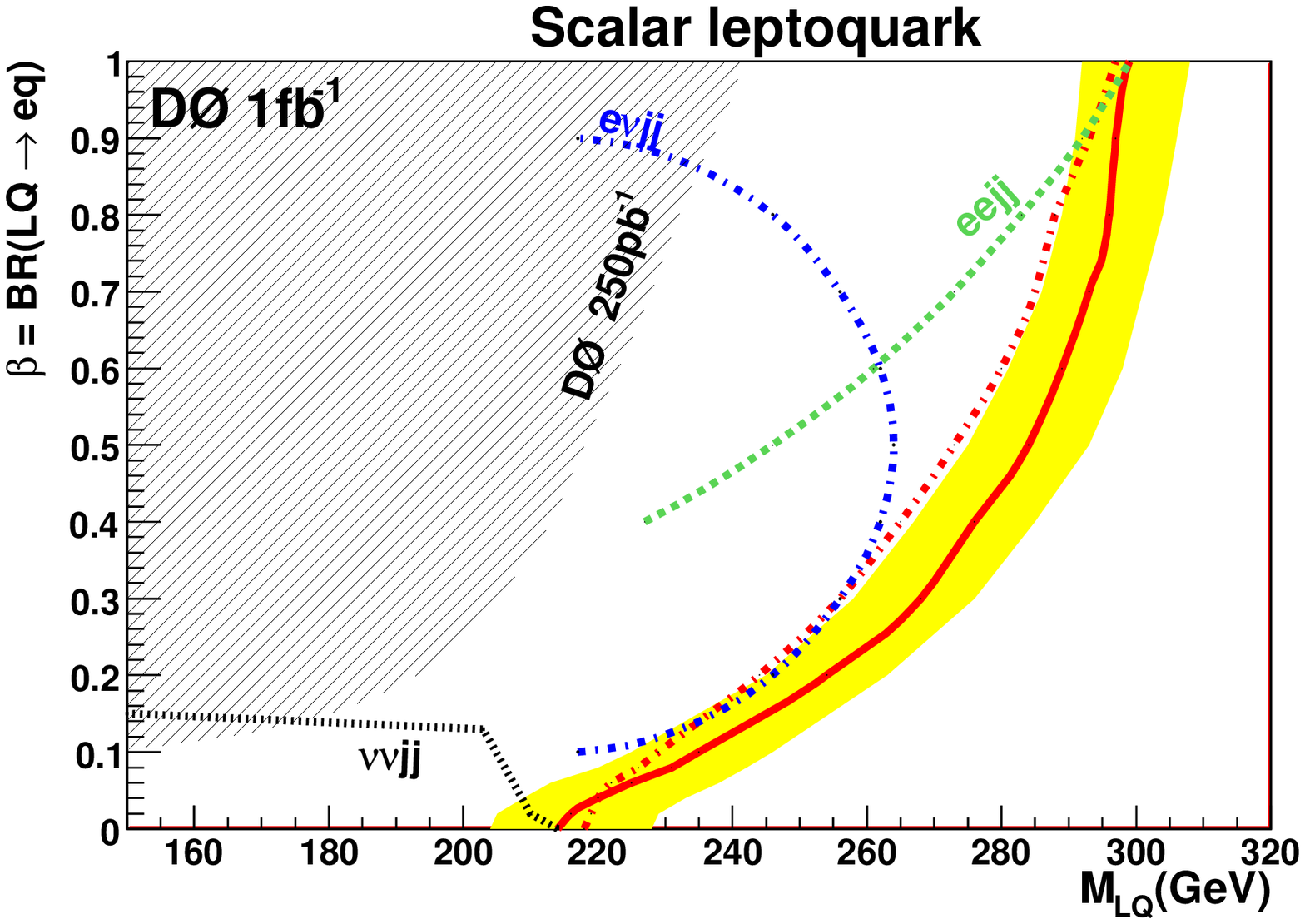}
\hspace{0.04\textwidth}\includegraphics[width=.45\textwidth,bb=0 14 514 438,clip]{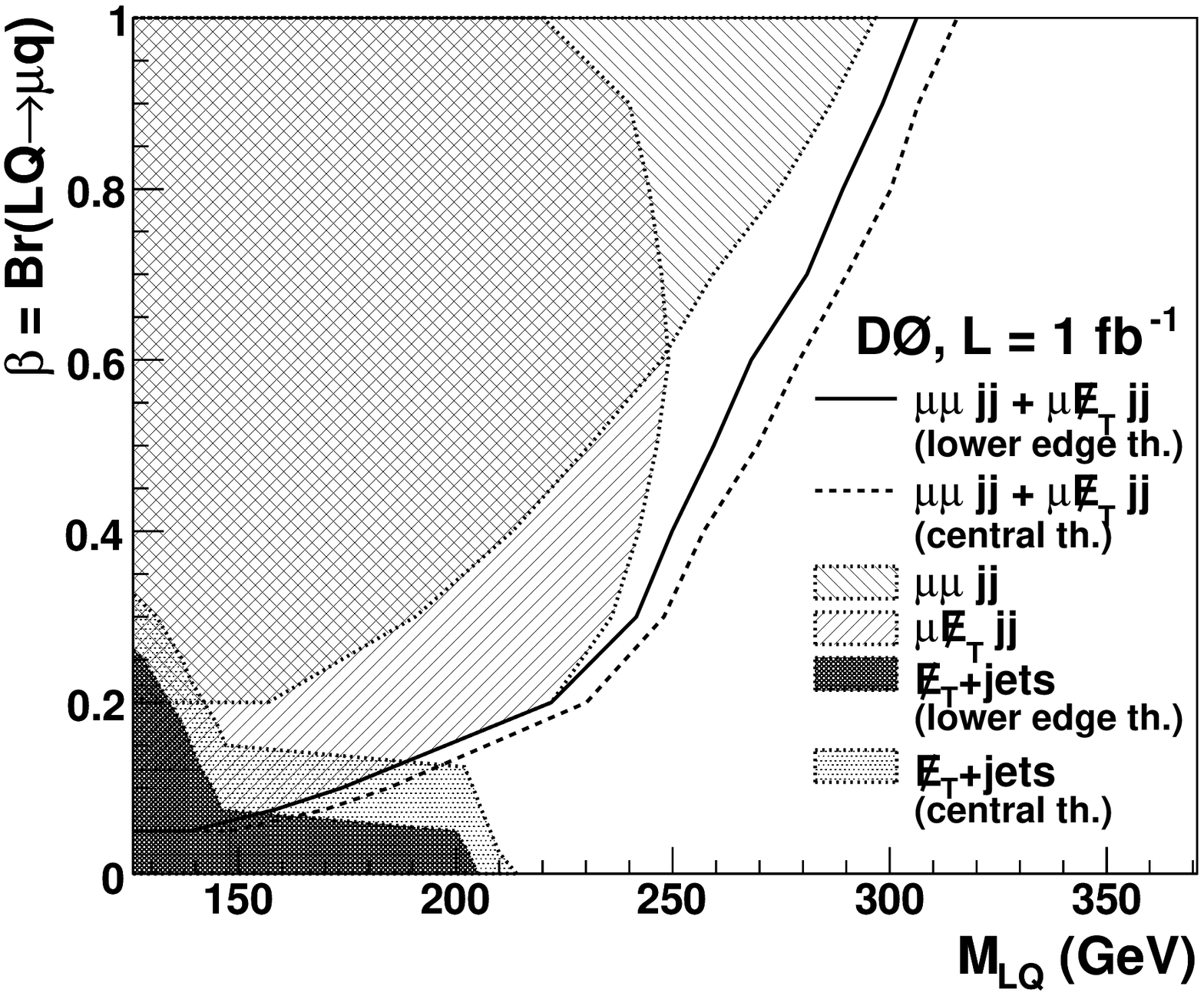}
\caption{The observed 95\% C.L. exclusion regions in the $M_{LQ}$ versus 
$\beta$ plane for first generation (left) and second generation (right)
scalar leptoquarks~\cite{Abazov:2009gf,Abazov:2008np}.}
\label{fig1}
\end{figure}

These two channels were combined with the search in the acoplanar jet topology
to derive an exclusion region for scalar first generation leptoquarks
in the $M_{LQ}$ versus $\beta$ plane (Fig.~\ref{fig1}, left). Limits on vector
leptoquarks were obtained as well, which generally reached higher mass scales.

\subsection{Second generation leptoquarks}
The search for the pair production of second generation scalar leptoquarks was
also recently published by the D0 collaboration using their $1\,\fbi$
data set~\cite{Abazov:2008np}.
A multivariate discrimination based on the {\em k-Nearest-Neighbors}\/
algorithm was employed: For both the $\mu j \mu j$ and $\mu j \MET j$ channels
six
kinematic input variables were used each (combinations of $S_T$, transverse
$\mu$-jet and $\nu$-jet masses, and $M_{\mu\mu}$). The main
systematic uncertainties were found to be due to the modeling of the
vector boson background, the jet energy scale, and the muon $p_T$ resolution.
The exclusion limit on $M_{LQ_2}$ as function of 
$\beta$ (Fig.~\ref{fig1}, right) reaches up to 316\,GeV at $\beta=1$.

\subsection{Third generation leptoquarks}
Searches for the pair-production of third generation leptoquarks were
performed in the $\tau b \tau b$ and $\nu b \nu b$ final states.

D0's search for $LQ_3\overline{LQ}_3\ra \tau b \tau b$ is based on the Run\,IIa
data set of $1\,\fbi$~\cite{Abazov:2008jp}. One of the taus was required to
decay into a muon ($\tau_\mu$) and the other tau needed to decay hadronically
($\tau_h$).
Hadronic $\tau$ decays were reconstructed from calorimeter clusters and tracks
and were separated into three types based on their decay.
The $\tau_h$ candidates as well as the $b$ quark jets were identified using
neural networks.

Lower limits on the scalar leptoquark mass were derived from the combination
of the single-tag and double-tag subsamples.
Assuming the hypothetical leptoquark has charge-$4/3$, which implies a 
branching fraction
$\mathcal{B}(LQ_3\ra \tau b)=1$, a lower mass limit $M_{LQ_3}>210\,\gev$
was set.
For charge-$2/3$ leptoquarks, decays into $\nu t$ are allowed as well, albeit
those are kinematically suppressed. Assuming equal leptoquark couplings to
$\tau b$ and $\nu t$, the mass limit only slightly decreases to 207\,GeV.
Based on the same final state, the CDF collaboration previously published
a search for the pair-produc\-tion of
third generation vector leptoquarks~\cite{Aaltonen:2007rb}. 

For the search $LQ_3\overline{LQ}_3\ra \nu b \nu b$ the D0 
collaboration presented a preliminary update based on $4\,\fbi$~\cite{d05931}.
The analysis required two or three reconstructed jets with one loose and
one tight $b$-tag and the final signal selection was based on $\MET$ and the
scalar sum of the jet transverse momenta.
Assuming that the leptoquarks have charge-{1/3} and 
that they decay exclusively in
a neutrino and a $b$ quark, a mass limit on third generation scalar leptoquarks
of $M_{LQ_3}>252\,\gev$ was derived.


\section{Search for quark compositeness in dijet angular distributions}
The angular distribution of dijets with respect to the beam direction directly
probes the dynamics of the underlying process. In the SM, the dijet cross 
section has a weak dependence on the variable
$\chi_{\mathrm{dijet}}= \exp(|y_1-y_2|)$, with $y_{1,2}$ 
being the jet rapidities. Quark compositeness (and numerous other new 
phenomena) 
would lead to an increased jet production rate at small 
$\chi_{\mathrm{dijet}}$ 
for large dijet masses $M_{jj}$. The D0 collaboration recently published a
measurement of the $\chi_{\mathrm{dijet}}$ distribution over a range of 
$M_{jj}$, from 0.25\,TeV to 1.1\,TeV, based on a data set corresponding to an
integrated luminosity of 0.7\,\fbi~\cite{Abazov:2009mh,parua}. Composite
quarks up to a scale of about 2.9\,TeV were excluded, improving previous
preliminary limits obtained by the CDF collaboration~\cite{cdf9609}.
In addition, bounds on the energy scale in various models with extra
spatial dimensions were derived.

\section{Conclusions}

The D0 and CDF experiments at the Tevatron collider searched for 
leptoquark production in a multitude of final states and for quark 
compositeness in dijet final states using data sets with 0.7 to 4\,\fbi.
The derived limits on leptoquark masses and compositeness scales improved
significantly compared to previous measurements. 
About $6\,\fbi$ of integrated luminosity has been recorded
by both experiments and they are expected to collect much larger data
sets during the full period of Run\,II.


\begin{thebibliography}{99}
\bibitem{lqpheno}
  J.~C.~Pati, A.~Salam, Phys.\ Rev.\ {D 10}, 275 (1974);
  H.~Georgi, S.~L.~Glashow, Phys.\ Rev.\ Lett.\ {32}, 438 (1974);
  W.~Buchm\"uller, D.~Wyler, Phys.\ Lett.\ {B 177}, 377 (1986).


\bibitem{Kramer:1997hh}
M.~Kramer, T.~Plehn, M.~Spira and P.~M. Zerwas,
Phys. Rev. Lett. {\bf 79} 341 (1997).

\bibitem{Abazov:2008at}
  V.~M.~Abazov {\it et al.}  [D0 Collaboration],
  Phys.\ Lett.\  B {\bf 668} 357 (2008).

\bibitem{CDFexoticweb}
  CDF Collaboration, {\tt http://www-cdf.fnal.gov/physics/exotic/}.

\bibitem{Abazov:2009gf}
  V.~M.~Abazov  [The D0 Collaboration],
  arXiv:0907.1048 [hep-ex].

\bibitem{Abazov:2008np}
  V.~M.~Abazov {\it et al.}  [D0 Collaboration],
  Phys.\ Lett.\  B {\bf 671} 224 (2009).

\bibitem{Abazov:2008jp}
  V.~M.~Abazov {\it et al.}  [D0 Collaboration],
  Phys.\ Rev.\ Lett.\  {\bf 101} 241802 (2008).

\bibitem{Aaltonen:2007rb}
  T.~Aaltonen {\it et al.}  [CDF Collaboration],
  Phys.\ Rev.\  D {\bf 77}, 091105 (2008).

\bibitem{d05931}
D0 Collaboration, Note 5931-CONF (2009).

\bibitem{Abazov:2009mh}
 V.~M.~Abazov {\it et al.}  [D0 Collaboration],
  arXiv:0906.4819 [hep-ex].

\bibitem{parua}
N. Parua, \emph{these proceedings}.

\bibitem{cdf9609} 
CDF Collaboration, Note 9609 (2008).

\end{thebibliography}
\end{document}